\documentclass[12pt,preprint]{aastex}

\def\kms  {km~s$^{-1}$}

\def\uas  {$\mu$as}

\def\etal {et al.~}
\def\eg   {e.g.~}
\def\ie   {i.e.~}
\def\hho  {H$_2$O}

\def\Ssrc {S~252}

\def\Vlsr {\ifmmode {V_{\rm LSR}}\else {$V_{\rm LSR}$}\fi}
\def\Ro   {\ifmmode {R_0}\else {$R_0$}\fi}
\def\To   {\ifmmode {\Theta_0}\else {$\Theta_0$}\fi}

\slugcomment{.}
\shorttitle{The Milky Way and Local Group} 
\shortauthors{Reid \etal}

\begin{document}

\title{Structure and Dynamics of the Milky Way: \\
       an Astro2010 Science White Paper}

\author{M. J. Reid (Harvard-Smithsonian CfA)}
\author{K. M. Menten (MPIfR), A. Brunthaler (MPIfR), G. A. Moellenbrock (NRAO) }
\author{}
\author{\epsscale{0.4}\plotone{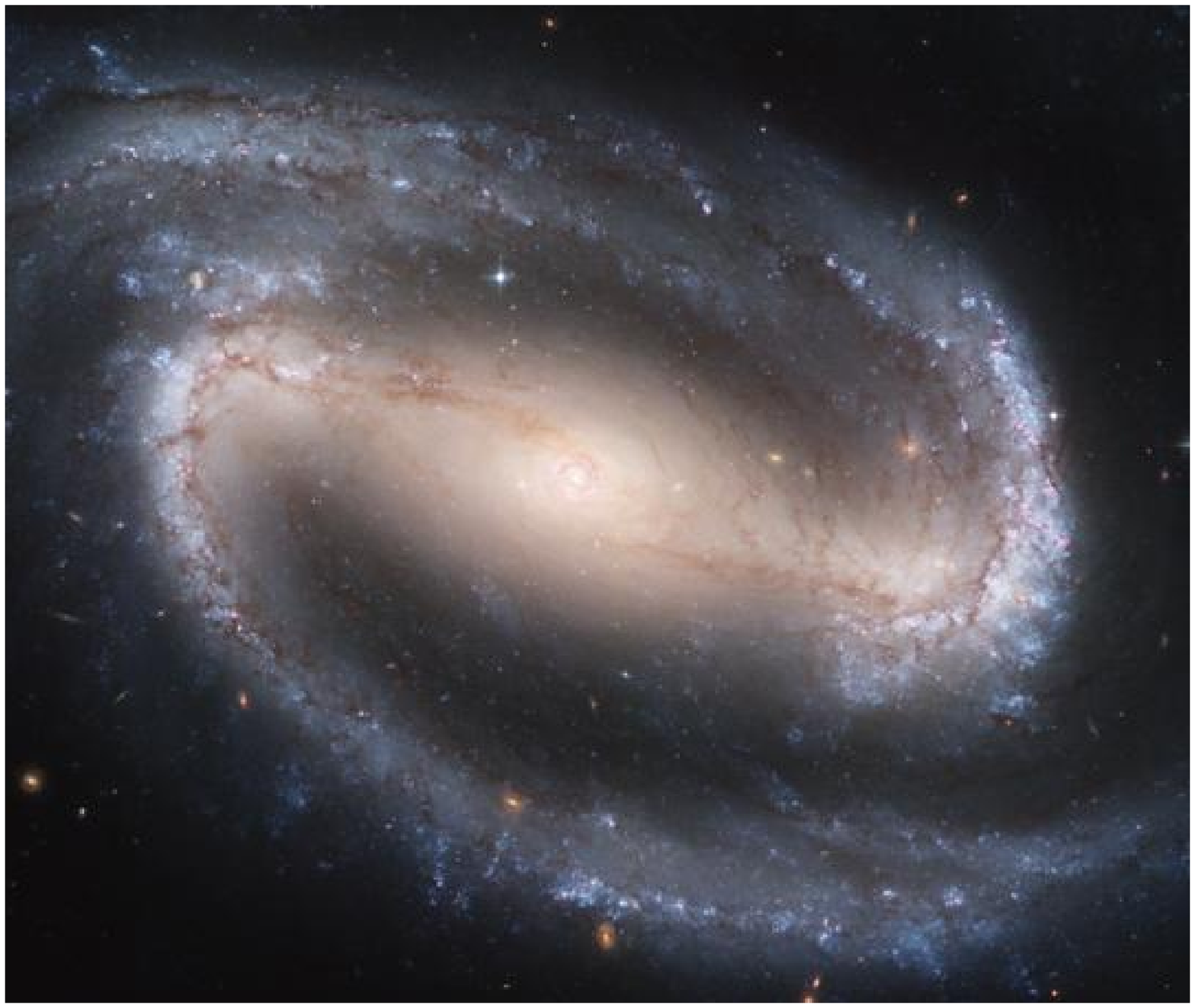}}
\author{NGC 1300       }
\author{\epsscale{0.4}\plotone{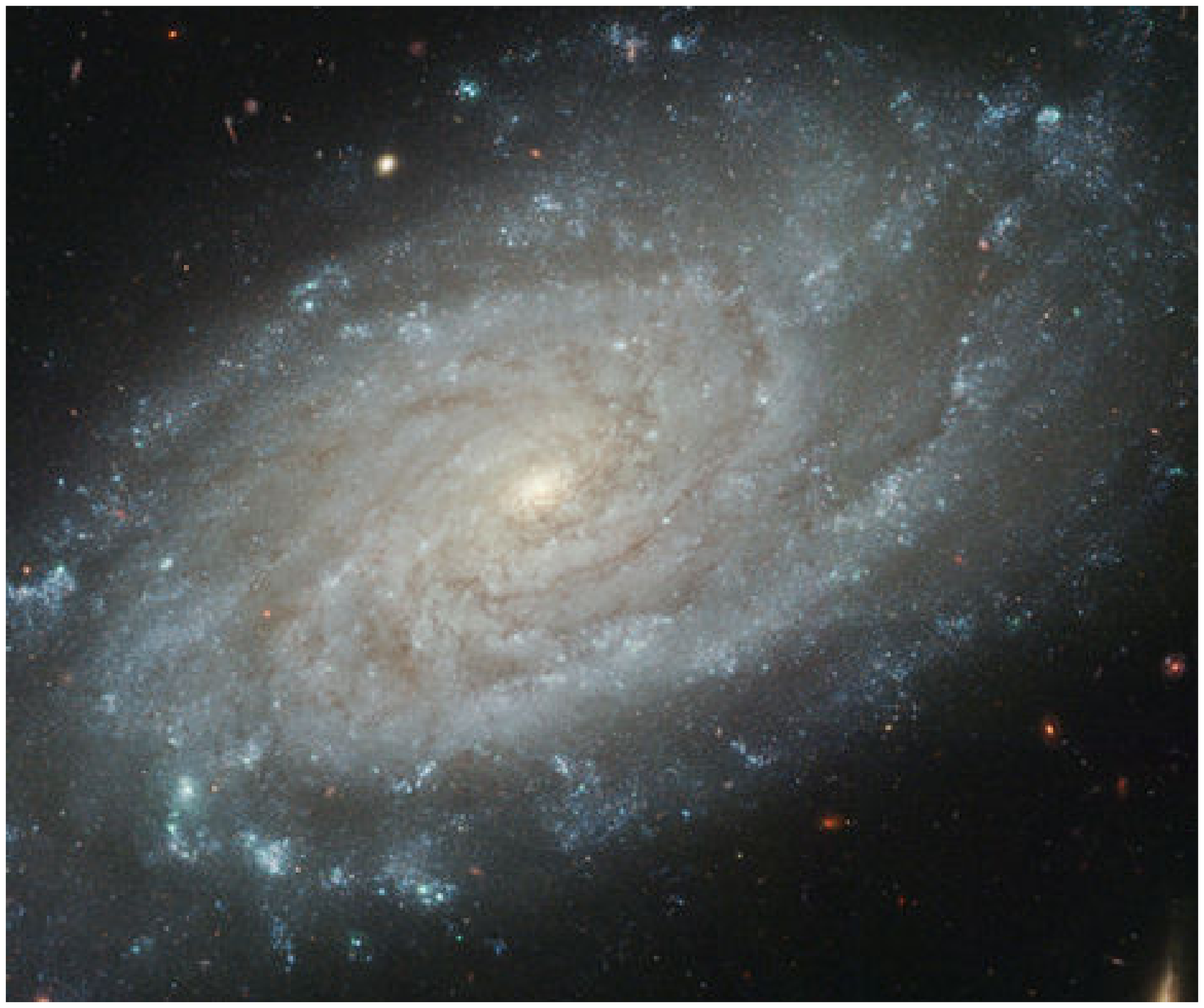}}
\author{NGC 3370       }
\author{       }
\author{\bf The Milky Way  ???     }

\begin{abstract}
Recent advances in radio astrometry with the VLBA have resulted in near micro-arcsecond  
accurate trigonometric parallax and proper motion measurements for masers in star forming 
regions.  We are now poised to directly measure the full 3-dimensional
locations and motions of {\it every} massive star forming region in the 
Milky Way and for the first time to map its spiral structure.  Such measurements would 
also yield the full kinematics of the Milky Way and determine its fundamental parameters 
(\Ro\ and \To) with 1\% accuracy.  Coupled with other observations this would yield the 
distribution of mass among the various components (including dark matter) of the Milky Way.
\end{abstract}

\section {Background}

Less than a century ago the very nature of ``spiral nebulae'' (galactic vs. 
extragalactic) was actively debated, whereas today we observe galaxies forming and 
interacting throughout the Universe.  Surprisingly, we know 
other galaxies far better than we know the Milky Way.  Since we are inside 
the Milky Way, it has proven very difficult to properly characterize its 
structure, because dust obscures most of the Galaxy at optical, and to some 
extent at IR, wavelengths and distances beyond the extended Solar Neighborhood are 
often quite uncertain.  Thus, we only have an ``educated guess'' that the 
Milky Way is a barred Sb or Sc galaxy, and even the number of spiral arms 
(2 or 4) is actively debated \citep{Benjamin:08}.

The discovery of a radio frequency transition of atomic hydrogen (HI at 21 cm
wavelength) in the 1950s offered the hope that, freed from extinction problems,
one could map the structure of the Milky Way.  
HI emission on Galactic longitude versus velocity plots clearly demonstrated 
that there are coherent, large-scale structures, which probably are spiral arms.  
However, determining accurate distances to HI clouds proved problematic, and 
this made the task of turning longitude-velocity data into a true ``plan-view'' of 
the Milky Way very uncertain \citep{Burton:88}.  Later, millimeter-wave observations 
of CO molecules also revealed coherent, large-scale structures with 
higher contrast than seen in HI \citep{Dame:01}.  But, again, uncertain 
distances to molecular clouds precluded making a true map of the Milky Way with 
sufficient accuracy to trace its spiral structure.

\begin{figure}[ht]
\epsscale{1.0}
\plottwo{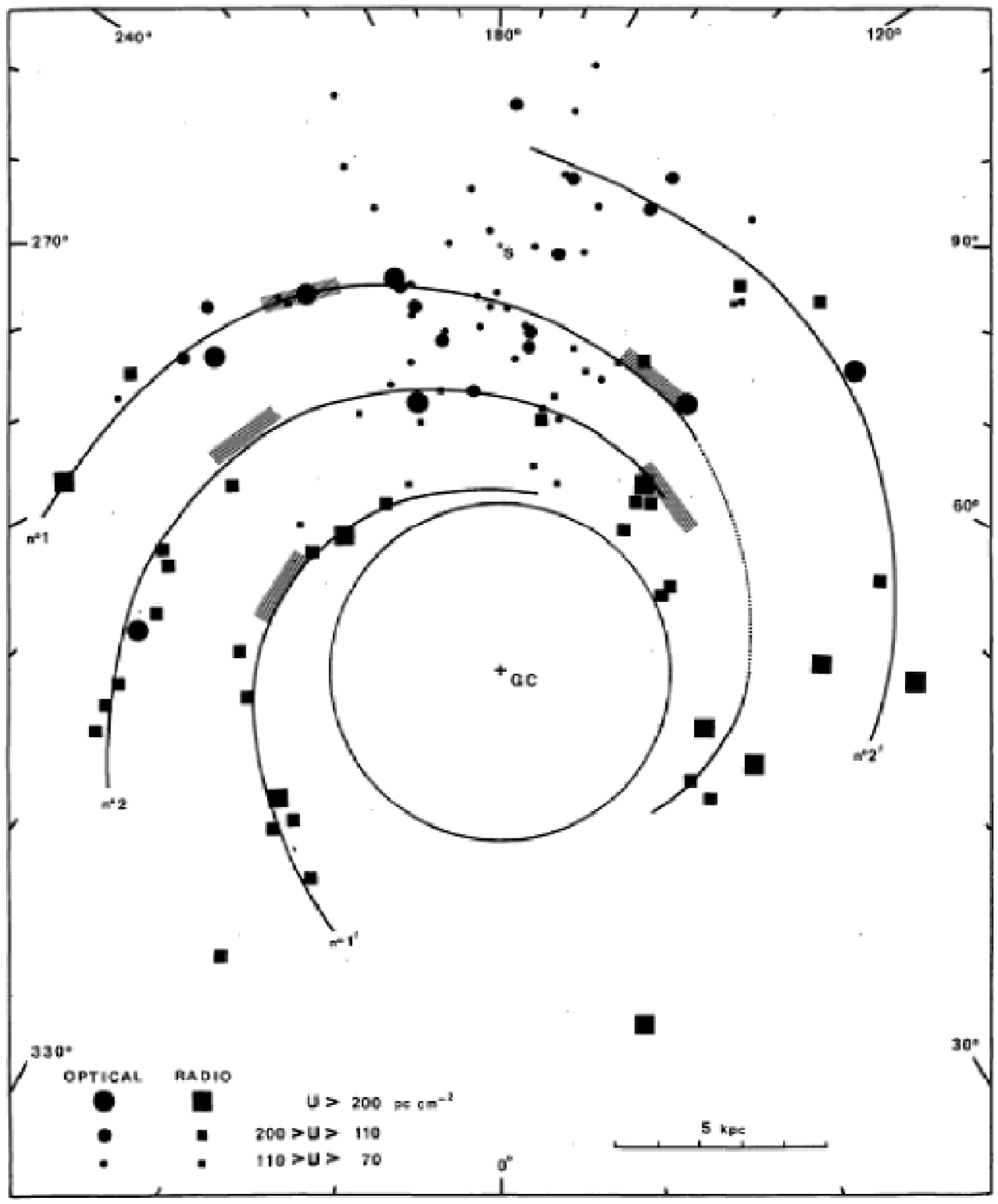}{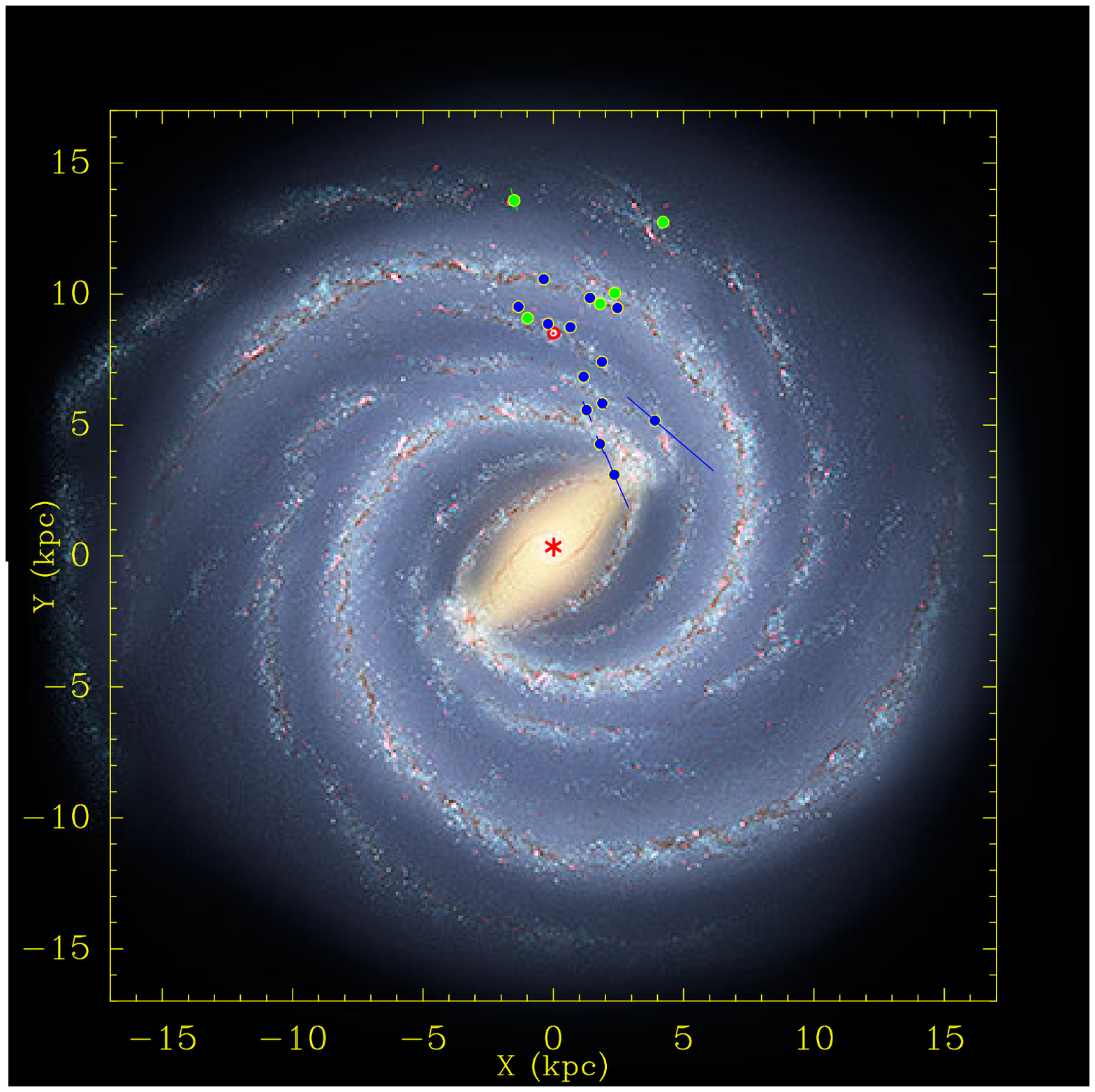} 
\caption{{
{\it Left Panel:} \citet{Georgelin:76} spiral model of HII regions 
in the Milky Way.  Considerable controversy exists as to the accuracy of 
this model, largely because many of the distances used are very uncertain.
Researchers in the field even disagree on the number of spiral arms.
Yet, over 30 years since publication, it remains the ``standard'' model.
{\it Right Panel:} Locations of high-mass star forming regions for which 
trigonometric parallaxes have been measured with VLBI.  Parallaxes of 
12 GHz methanol masers are indicated with {\it dark blue dots} and those
from water masers are indicated with {\it light green dots}.
Distance error bars are indicated, but most are smaller than the dots.
The Galactic center ({\it red asterisk}) is at (0,0) and the Sun 
({\it red Sun symbol}) at (0,8.5).
The background is an artist's conception of Milky Way  
(R. Hurt: NASA/JPL-Caltech/SSC) viewed from the NGP.  
The artist's image has been scaled to place 
the star forming regions in the spiral arms.
         }}
\label{fig:milky_way}
\end{figure}

\citet{Georgelin:76} constructed a plan-view model of the spiral structure of 
the Milky Way (see Fig.~\ref{fig:milky_way}).  Their approach 
involved combining optical observations of young stars and radio data of HI 
clouds and HII regions.  Luminosity distances to nearby stars were 
used where available and kinematic distances elsewhere, mostly for distant 
HII regions.  While subject to very significant uncertainties from kinematic 
distances, the Georgelin model has remained the ``standard'' model of the spiral 
structure of the Milky Way for over 30 years.  However, debate continues
over such basic facts as the existence of some spiral
arms, the number of arms, and the size, rotation speed and mass of the Milky Way. 

\section {Scientific Context: Mapping the Milky Way}

\begin{figure}[ht]
\epsscale{0.75}
\plotone{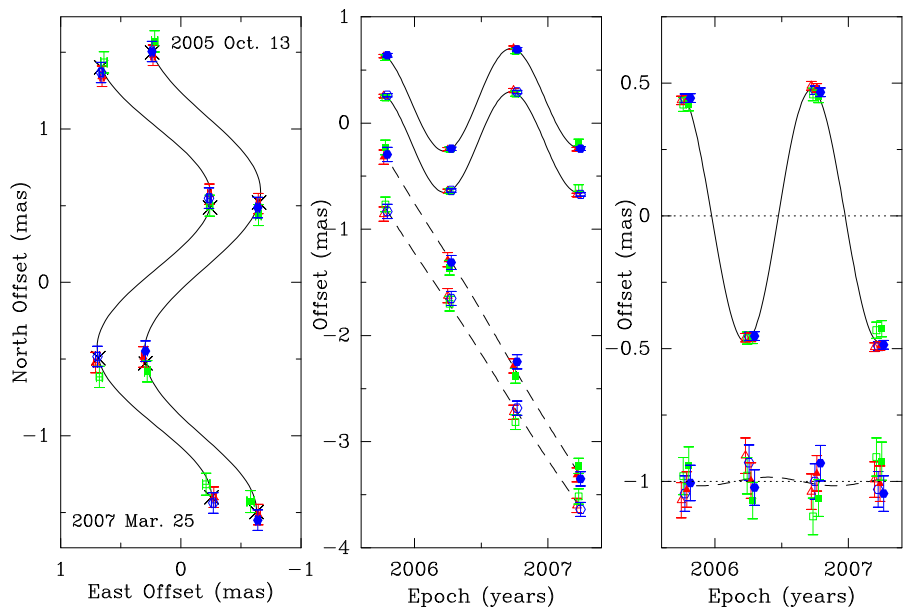} 
\caption{{
  Astrometric data for \Ssrc\ showing the parallax fit of $476\pm6$~\uas\ 
  from \citet{Reid:09a}.
  Plotted are position measurements of two maser spots {\it (open and solid symbols)} 
  relative to the three background quasars: J0603+2159 {\it (red triangles)}, 
  J0607+2218 {\it (green squares)} and J0608+2229 {\it (blue hexagons)}.
  {\it Left Panel:} Positions on the sky with first and last epochs labeled.
  Data for the two maser spots are offset horizontally for clarity.
  The expected positions from the parallax and proper motion fit
  are indicated {\it (crosses)}.
  {\it Middle Panel:} East {\it (solid lines)} and North {\it (dashed lines)} position
  offsets and parallax and proper motions fits versus time.  
  Data for the two maser spots are offset vertically, the northward
  data have been offset from the eastward data, 
  and small time shifts have been added to the data for clarity.
  {\it Right Panel:} Same as the {\it middle panel}, except the
  best-fit proper motions have been removed, allowing
  all data to be overlaid and the effects of only the parallax seen.
        }}
\label{fig:s252_parallax}
\end{figure}

Recent improvements in radio astrometry with the VLBA have 
yielded parallaxes and proper motions to star forming regions across a 
significant portion of the Milky Way with accuracies of $\sim10$~\uas\ and 
$\sim1$~\kms, respectively 
\citep{Reid:09a,Moscadelli:09,Xu:09,Zhang:09,Brunthaler:09,Moellenbrock:09}.
Fig.~\ref{fig:s252_parallax} shows the
results of VLBA observations of 12 GHz methanol masers associated with
a massive young stellar object in the star forming region S~252.
The data yield a parallax of $476\pm6$~\uas.
With such data one can measure distances to sources on the far side
of the Milky Way with 10\% accuracy!  

Combining the results from 16 similar measurements shows the potential
of VLBA parallaxes for mapping the Milky Way \citep{Reid:09b}.  The results begin 
to locate spiral arms (see Fig.~\ref{fig:milky_way}) and yield the first direct 
measurement of arm pitch angles.  In addition, estimates of the
fundamental parameters of the Milky Way, \Ro\ and \To, indicate a
rotation speed of $\To=254$~\kms, some 15\% faster than usually assumed.
Interestingly, these first astrometric results indicate that the Milky Way 
and the Andromeda galaxy \citep{Carignan:06} have nearly identical rotational 
properties, suggesting similar dark matter sizes and masses and contrary
to the general assertion that Andromeda is significantly more massive than
the Milky Way.

Changing the value of \To\ from the IAU standard 220~\kms\ to $\approx250$~\kms\ significantly
affects models of the Local Group of galaxies.   It results in a decrease of about
20~\kms\ in the space velocity of the LMC {\it relative to the center of the Milky Way} and an
increase of about 50\% in the estimated (dark matter) mass of the Milky Way.  
Both help to bind the LMC to the Milky Way \citep{Shattow:08} and reverse the
conclusion, based on HST measurements of the proper motion of the LMC,
that the LMC was unbound and making its first pass near the Milky Way 
\citep{Kallivayalil:06}.

We are currently poised to make truly dramatic progress in understanding 
the Milky Way.  Over the next 10 to 20 years, we could measure the distance
to {\it every} high mass star forming region in the Galaxy with parallax accuracies
of $\sim1$~\uas.   At the far side of the Galaxy (16~kpc) this would correspond to
better than $\sim2$\% distance accuracy.  {\it Thus, for the first time we could map in
detail the spiral structure of the Milky Way and learn what it really looks like.}  

Indeed, with this accuracy one could easily resolve structure, not only across
spiral arms, but also across the bar in the Galactic center region, provided the
density of target sources is adequate to reveal structure.
Finally, we would have not only the 3-dimensional locations of all major star forming regions, 
but also their 3-dimensional velocity vectors.  This would yield extraordinarily
accurate measurements (projected to be better than $\pm1$\%) of such fundamental parameters 
as the distance to the Galactic center (\Ro), the rotation speed 
of the LSR (\To), the form of the rotation curve, and the kinematic effects 
of spiral structure.   
Note that GAIA or SIM, which may have comparable astrometric accuracy,
will operate at optical wavelengths and cannot see through the dust in the
plane of the Milky Way, nor to deeply embedded regions of star formation.
Since the ionizing radiation from OB-type stars in high mass star forming regions
in the Galactic plane best defines spiral structure, the most straightforward way 
to map this structure is through radio astrometry.

\section {Telescope Needs}

\begin{deluxetable}{ll}
\tablecolumns{2} \tablewidth{0pc} 
\tablecaption{Telescopes Advances and their Scientific Impact}
\tablehead {
  \colhead{Telescope Advance} & \colhead{Scientific Impact} 
            }
\startdata
6.7 GHz receivers for VLBA &
Expand number of parallax targets by $\times10$ \\
& to map spiral structure and dynamics of the \\
& (northern) Milky Way\\
\\
High (32 Gbps) data recording rate and/or 
&
Parallax calibrators a factor of $>6$ nearer  \\
additional telescopes/collecting area  
& to targets, enabling sub-\uas\ astrometry \\
\\
Improved southern hemisphere VLBI  &
Map southern portion of Milky Way\\
capability (\eg partial SKA) & \\
\enddata
\label{table:telescope_impact}
\end{deluxetable}

The observations needed to map the Milky Way outlined above can be accomplished with the
advances outlined in Table~\ref{table:telescope_impact}.  
Some of the goals can be achieved with modest upgrades in receiver and data 
recording equipment at the VLBA.  Currently most of the parallax measurements
with the VLBA have employed 12 GHz methanol masers.
Methanol masers associated with high mass star forming regions are nearly
ideal astrometric targets; they are compact, long-lived, and their motions
are closely tied to the massive star that excites them.  (\hho\ masers are not
a good substitute.  They participate in fast outflows 
and have lifetimes less than the 1 year necessary for good parallax measurements.
Also, their large internal motions make it difficult to associate a maser  
motion with that of its central massive star.  This latter problem does not affect parallax 
measurement, but it does limit the interpretation of the proper motions for Galactic dynamics.)
Unfortunately, there are only several tens of 12 GHz methanol masers that
are strong enough for VLBA parallax measurement.   Adding a new VLBA receiver, capable
of observing the much stronger 6.7 GHz methanol masers, is needed to map the 
locations and motions of hundreds of star forming regions 
across major portions of the Milky Way.  

Upgrading the VLBA data recording rate by more than two orders of magnitude 
from 256 Mbps to 32 Gbps (which requires no new technology) would dramatically 
improve astrometric accuracy by making far more background quasars available as 
position references.  VLBA astrometric observations are usually limited by systematics 
that cancel proportionally to the separation of the maser target and background quasar.
The factor of 11 (\ie $\sqrt{32 {\rm Gbps}/256 {\rm Mbps}}$) improvement in continuum 
sensitivity from the increased recording capability would lead to an average decrease in 
target-quasar separation by a factor of 6 (assuming standard $\log{N}/\log{S}$
statistics).  This should allow parallaxes accurate 
to better than $\pm1$~\uas!

In order to map the entire Milky Way, we will need a VLBA-like capability
in the southern hemisphere.   This could be well met by the ``SKA-mid'' project,
which will be placed in the southern hemisphere, provided the antennas 
can reach the 6.7 GHz transition of methanol masers.  Alternatively
a relatively modest upgrade to the Australian VLBI capabilities could also
provide the required capabilities.

Combining the above mentioned advances in receiver and recording capabilities
with increased collecting area would allow measurement of weaker target sources.   
This could be achieved in conjunction with a ``path-finder'' project that would 
prototype and test antenna ``patches'' with $\sim5$\% of an SKA collecting area.  
Placing some of these antenna patches at VLBA sites and at some new sites
between the EVLA and the VLBA would greatly increase the sensitivity of the array.  
Some of these telescope advances could come from the phased 
implementation of the plans outlined in the ``North American Array'' initiative 
(J. Ulvestad, coordinator) submitted to the Decadal Survey.  The
construction of even a 5\% SKA in the Southern Hemisphere would allow exquisite 
mapping of the entire Milky Way.  Finally, we note that the construction of the 
entire SKA concept would revolutionize all of these activities and lead to truly 
dramatic astrometric results.


\begin{thebibliography}{}
\bibitem[Benjamin (2008)]{Benjamin:08} Benjamin, R. A. 2008,
    in Massive Star Formation: Observations Confront Theory,
    eds. H. Beuther, H. Linz \& Th. Henning,
    ASP Conference Series, Vol. 387, p. 375
\bibitem[Brunthaler \etal (2009)]{Brunthaler:09} 
    Brunthaler, A., Reid, M. J., Menten, K. M., Zheng, X. W., Moscadelli, L. \&
    Xu, Y.  2009, arXiv:0811.0713, to appear in \apj, 692, 
\bibitem[Burton (1988)]{Burton:88} Burton, W. B. 1988, in Galactic and Extragalactic
    Radio Astronomy, 2nd ed., eds. G. L. Verschuur \& K. I. Kellermann,
    (Springer-Verlag, New York), p. 295
\bibitem[Carignan \etal (2006)]{Carignan:06} Carignan, C., Chemin, L., Huchtmeier, W. K.
    \& Lockman, F. J. 2006, \apj, 641, L109    
\bibitem[Dame, Hartmann \& Thaddeus (2001)]{Dame:01} Dame, T. M., Hartmann, D. 
    \& Thaddeus, P. 2001, \apj, 547, 792
\bibitem[Georgelin \& Georgelin (1976)]{Georgelin:76} Georgelin, Y. M. \& 
    Georgelin, Y. P., 1976, \aap, 49, 57,
\bibitem[Kallivayalil \etal (2006)]{Kallivayalil:06}
    Kallivayalil, N., van der Marel, R. P., Alcock, C., Axelrod, T., Cook, K. H.,
    Drake, A. J. \& Geha, M. 2006, \apj, 638, 772
\bibitem[Moellenbrock, Claussen \& Goss (2009)]{Moellenbrock:09} Moellenbrock, G. A., 
    Claussen, M. J. \& Goss, W. M. 2009, arXiv:0901.0517, to appear in \apj, 694
\bibitem[Moscadelli \etal (2009)]{Moscadelli:09} 
    Moscadelli, L, Reid, M. J., Menten, K. M., Brunthaler, A., Zheng, X. W. \&
    Xu, Y.  2009, arXiv:0811.0679, to appear in \apj, 692, 
\bibitem[Reid \etal (2009a)]{Reid:09a} 
    Reid, M. J., Menten, K. M., Brunthaler, A., Zheng, X. W., Moscadelli, L. \&
    Xu, Y.  2009a, arXiv:0811., to appear in \apj, 692, 
\bibitem[Reid \etal (2009b)]{Reid:09b} 
    Reid, M. J. \etal 2009b, to be submitted to \apj (check Astro-ph shortly) 
\bibitem[Shattow \& Loeb (2008)]{Shattow:08} Shattow, G. \& Loeb, A. 2008, 
    \mnras, 392, L21
\bibitem[Xu \etal (2009)]{Xu:09} 
    Xu, Y, Reid, M. J., Menten, K. M., Brunthaler, A., Zheng, X. W. \& Moscadelli, L.
    2009, arXiv:0811.0701, to appear in \apj, 692, 
\bibitem[Zhang \etal (2009)]{Zhang:09} 
    Zhang, B., Zheng, X. W., Reid, M. J., Menten, K. M., Xu, Y., Moscadelli, L. \& 
    Brunthaler, A. 2009, arXiv:0811.0704, to appear in \apj, 692, 
\end{thebibliography}
\end{document}